\documentclass[fleqn,usenatbib]{mnras}

\usepackage{newtxtext,newtxmath}

\usepackage[T1]{fontenc}

\DeclareRobustCommand{\VAN}[3]{#2}
\let\VANthebibliography\thebibliography
\def\thebibliography{\DeclareRobustCommand{\VAN}[3]{##3}\VANthebibliography}

\usepackage{graphicx}	
\usepackage{amsmath}	

\newcommand{\hessj}{HESS\,J1832$-$093}
\newcommand{\ircounter}{2MASS\,J18324516$-$0921545}

\title[Spectral classification of HESS J1832-093]{NIR spectral classification of the companion in the gamma-ray binary HESS J1832--093 as an O6 V star}

\author[B. van Soelen et al.]{
B. van Soelen,$^{1}$\thanks{E-mail: vansoelenb@ufs.ac.za}
P. Bordas,$^{2}$\thanks{Email: pbordas@fqa.ub.edu}
I. Negueruela,$^{3}$
E. de O\~na Wilhelmi,$^{4}$
A. Papitto$^{5}$
and M. Rib\'o$^{2}$\thanks{Serra H\'unter Fellow}
\\
$^{1}$Department of Physics, University of the Free State, PO Box 339, Bloemfontein 9300, South Africa\\
$^{2}$Departament de Física Quàntica i Astrofísica, Institut de Ciències del Cosmos, Universitat de Barcelona, IEEC-UB, Martí i Franquès 1, 08028, Barcelona, Spain\\
$^{3}$Departamento de Física Aplicada, Facultad de Ciencias, Universidad de Alicante, Carretera de San Vicente s/n, 03690 San Vicente del Raspeig, Spain\\
$^{4}$ Deutsches Elektronen-Synchrotron DESY, Platanenallee 6, 15738 Zeuthen, Germany \\
$^{5}$ INAF Osservatorio Astronomico di Roma, via Frascati 33, Monteporzio Catone, Roma, I-00078, Italy
}

\date{Accepted XXX. Received YYY; in original form ZZZ}

\pubyear{2023}

\begin{document}
\label{firstpage}
\pagerange{\pageref{firstpage}--\pageref{lastpage}}
\maketitle

\begin{abstract}
\hessj\ is a member of the rare class of gamma-ray binaries, as recently confirmed by the detection of orbitally modulated X-ray and gamma-ray emission with a period of $\sim$86 d. The spectral type of the massive companion star has been difficult to retrieve as there is no optical counterpart, but the system is coincident with a near-infrared source. Previous results have shown that the infrared counterpart is consistent with an O or B type star, but a clear classification is still lacking. We observed the counterpart twice, in 2019 and 2021, with the X-Shooter spectrograph operating on the VLT. The obtained spectra classify the counterpart as an O6~V type star. We estimate a distance to the source of $6.7 \pm 0.5$\,kpc, although this estimate can be severely affected by the high extinction towards the source. This new O6~V classification for the companion star in \hessj\ provides further support to an apparent  grouping around a given spectral type for all discovered gamma-ray binaries that contain an O-type star. This may be due to the interplay between the initial mass function and the wind-momentum-luminosity relation. 

\end{abstract}




\section{Introduction}

The production of very-high-energy (VHE) gamma-ray emission from astrophysical sources requires both a powerful particle accelerator and the existence of favourable ambient conditions upon which these particles can interact. The latest generation of Cherenkov telescopes have imaged about a hundred such TeV emitters, the vast majority of which are located in the Galactic Plane \citep{hess_galactic_plane2018}. While VHE emission from a variety of Galactic sources has now been revealed (including pulsars and pulsar wind nebulae, gamma-ray binaries, supernova remnants, and stellar clusters) so far gamma-ray binaries are, together with pulsars and novae, the only Galactic cases in which  variable TeV emission has been reported. {Gamma-ray binaries are high mass binary systems which consist of a neutron star or black hole, which orbits an O or Be type companion, and which produce persistent non-thermal emission which peaks (in a $\nu F_\nu$ distribution) in the gamma-ray regime \citep[e.g.][]{Dubus13}. Gamma-ray binaries represent a unique framework to study the mechanisms through which VHE emission is produced, as they provide variable photon- and matter-field ambient conditions in the emitter. Observations of gamma-ray binaries can render key information on the particle acceleration mechanisms and radiation/absorption processes taking place in the close vicinity of a compact object. 

So far,  only nine gamma-ray binaries have been identified, namely PSR\,B1259$-$63, LS\,5039, LS\,I\,+61\,303, HESS\,J0632+057, \hessj,  1FGL\,J1018.6$-$5856, LMC P3, PSR\,J2032+4127, and 4FGL J1405.1$-$6119  \citep[][respectively]{hess05, aharonian05_ls5039_tev_detection, albert06, aharonian07, HESS15b, HESS_1FGL1018, corbet16, veritas_magic_j2032_2018, Corbet19}.\footnote{The source HESS~J1828$-$099 has been recently proposed as a gamma-ray binary candidate, see \citet{DeSarkar2022}}
All of them are high-mass systems composed of a compact object in the mass range of a black hole or a neutron star orbiting a luminous O or Be type companion star, and exhibit the maximum of their non-thermal emission at gamma-ray energies. In only three of these systems, namely PSR~B1259$-$63, PSR J2032+4127 and LS\,I\,+61\,303 \citep{Johnston92, Camilo09,  Weng2022}, is the nature of the compact object known due to the detection of pulsed emission.\footnote{An indication of pulsed emission has also been reported for LS\,5039 by \citet{yoneda20,makishima23}; but see also \citet{volkov21,kargaltsev23}.} While there are similarities between the gamma-ray binaries,  each source shows its own distinct characteristics in its spectral properties and phase folded flux profile. However, there is a general trend that the gamma-ray binary systems that contain O-type stars show more regular orbit-to-orbit behaviour \cite[e.g.\ LS\,5039;][]{mariaud15}, while those that contain Be-type stars have shown super-orbital periods  \citep[LS\,I +61 303;][and references therein]{ahnen16}, local maxima in their light curves associated with the compact object crossing the circumstellar disc \cite[e.g.\ HESS\,J0632+057, PSR\,B1259$-$63, PSR\,J2032+4127;][]{moritani18,hess05,abeysekara18}, as well as significant orbit to orbit variability \citep[e.g.\ PSR\,B1259$-$63][]{chernyakova21}. There also seems to be a general trend where O-type systems have shorter orbital periods than Be-type systems \citep[see e.g.\ table~1 in][]{chernyakova19}. Deeper studies are required to understand whether there exists a unified physical picture describing all of them as a class.

\hessj\ was serendipitously discovered close to the rim of SNR G22.7$-$0.2 during observations of the Galactic plane with the H.E.S.S. telescopes \citep{HESS15b}. The source displays a differential TeV flux of ($4.8 \pm 1.8) \times 10^{-13}$~TeV$^{-1}$~cm$^{-2}$~s$^{-1}$ with a spectral index $\Gamma_{\gamma} = 2.6 \pm 0.4$, and appears point-like at TeV energies.   No statistically significant variability was found in the TeV data, and its identification remained uncertain. Three possibilities were presented given its point-like nature: a gamma-ray binary scenario, a young pulsar wind nebula, or a background Active Galactic Nucleus  \citep{HESS15b}.

A search for multi-wavelength counterparts was conducted to further constrain the system properties. Observations with XMM-{\it{Newton}} in 2011 revealed a relatively faint ($\phi_{\rm 2-10\,keV}  = 6.9_{-2.8}^{+1.7} \times 10^{-13}$~erg~cm$^{-2}$~s$^{-1}$), hard ($\Gamma_{\rm X} = 1.3_{-0.4}^{+0.5}$) and highly absorbed ($N_{\rm H} = 10.5_{-2.7}^{+3.1} \times 10^{22}$~cm$^{-2}$) point-like X-ray source coincident with the position of the TeV source \citep{HESS15b}. In addition, an infrared counterpart, 2MASS\,J18324516--0921545 \citep[][apparent magnitudes $J = 15.52 \pm 0.06$~mag, $H = 13.26 \pm 0.04$~mag, and  $K_S = 12.17 \pm 0.02$~mag]{Skrutskie2006}, was found $\sim 1.9\arcsec$ away from the best fit XMM-{\it{Newton}} position. The chance probability of such a spatial coincidence was found to be $\lesssim$ 2 per cent, prompting an association between \hessj\ and 2MASS\,J18324516--0921545 (\citealt{HESS15b}).

Further X-ray observations strengthened the case for a gamma-ray binary. \citet{Eger2016} reported that observations of \hessj\ with {\it Chandra} in 2015 displayed a 2--10 keV flux $\sim$6 times higher than the one obtained with XMM-{\it{Newton}} in 2011, with the spectral parameters remaining essentially unchanged. Note, however, that a smaller change in the flux was found in a re-analysis by \citet{Mori2017}. Additionally, \citet{Eger2016} placed  
a limit on the pulsed fraction at the level of $\sim$ 45 per cent. These observations also refined the X-ray source position to within  0.3$\arcsec$ of its proposed infrared counterpart.  Further {\it NuSTAR} observations of the source in 2016 showed that the X-ray spectrum is well fitted with a power-law ($\Gamma = 1.5$) up to 30 keV, with no indication of a break \citep{Mori2017}. This is consistent with gamma-ray binaries that do not show the characteristic break of High Mass X-ray Binaries (HMXBs) at keV energies \citep[e.g.][]{Dubus13}. \citet{Mori2017} also undertook timing analysis which showed no evidence of pulsation or accretion {signatures}. In the GeV domain, \hessj\ has been detected with the \textit{Fermi}-LAT by \citet{marti20}. Critically, these authors reported on the discovery of  orbitally modulated emission at both X-ray and gamma-ray energies, with a period of $\sim 86.3$\,d in the {\it Swift} data, and $\sim 87.0$\,d in the {\it Fermi}-LAT data. This detection confirmed that this source is a gamma-ray binary.

No counterpart to \hessj\ has been identified at optical wavelengths. \citet{Mori2017}, found that the absolute J and K magnitudes for the near-infrared (NIR) counterpart would be compatible with a B8V or B1.5V star, based on a hydrogen column density of $N_{\rm H}=1.7\times10^{22}$\,cm$^{-2}$ from radio surveys and a distance of 4.4\,kpc, but cautioned that the higher column density for X-ray observations suggests this is a lower limit and higher extinction would imply an O-type star. More recently, \citet{Tam2020} reported on \textit{Gemini} NIR spectroscopic observations which found the counterpart to be consistent with a late O-type or early B-type star.

In this letter we report on new upper-limits on the optical magnitude of the optical counterpart, as well as new near-infrared spectroscopic observations undertaken with the X-shooter spectrograph operating on the VLT in Paranal, Chile, intended to definitively classify the spectral type of the proposed counterpart.

\section{Observations of \hessj}
\label{section:observations}
\subsection{TJO and NOT observations}
\label{sec:NOT}

Observations were undertaken on 2015 October 31 to search for the optical counterpart of 2MASS 18324516--0921545 (RA=18:32:45.162s, DEC=$-$09:21:54.55) in the R-band using MEIA2 on the fully robotic 0.8-m Joan Or\'o telescope (TJO) at the Montsec Observatory (OdM) \citep{2013hsa7.conf..958V}. MEIA2 is a 2048$\times$2048 back-illuminated CCD detector achieving a 12.3\arcmin$\times$12.3\arcmin\ field of view. No optical counterpart could be found in the R-image in the 900\,s observations. The closest sources, with a magnitude of 19.3, are located at 5.6$^{\prime\prime}$ and 8.2$^{\prime\prime}$ (calibrated with stars from the USNO B-1.0 catalogue, yielding an astrometric RMS uncertainty of 0.4$^{\prime\prime}$) from the \textit{Chandra} and 2MASS source (astrometric precision of 20 mas). Follow-up imaging observations using the 2.65-m  Nordic Optical Telescope (NOT) telescope were obtained (proposal 118-Multiple-2$\/$16A) with ALFOSC (2016 March 22) and NOTCam (2016 August 12). Optical observations were taken with ALFOSC with a 2048$\times$2064 CCD detector achieving a 6.4\arcmin$\times$6.4\arcmin field of view, while infrared observations were taken with NOTCam which has a 1024$\times$1024 detector which achieves a 4\arcmin$\times$4\arcmin field of view.  The optical observations constrained the magnitude of the counterpart to be fainter than $B\sim21.1$, $R \sim 22.3$, and $I \sim 21.8$\,mag (3$\sigma$ CL), while the infrared observations found $J = 15.54 \pm 0.01$\,mag, consistent with the 2MASS catalogue.

\subsection{X-shooter observations and data reduction}

The  infrared counterpart associated with \hessj, \ircounter, was observed twice using X-Shooter \citep{Vernet2011} on the VLT, on 2019 October 4 and 2021 June 28.  X-Shooter, mounted at the Cassegrain focus of the 8.2-m UT3 telescope, consists of three echelle spectrographs which cover the ultraviolet, optical and NIR wavelength range. Both observations were undertaken in `AutoNodOnSlit' mode, and consisted of a 9$\times$175\,s exposure of the target using only the NIR detector, with the $0.9\arcsec$ slit ($R\sim 5600$). The first observation achieved a signal-to-noise ratio of $\sim$136 and $\sim$138 in the H and K bands respectively, while the second observation achieved $\sim$98 and $\sim$102. Both observations were undertaken with seeing $<$1\arcsec. Data reduction and spectral extraction was performed with the ESO XShooter pipeline \citep{Modigliani10}. Telluric correction was performed using {\sc molecfit} \citep{Smette2015, Kausch2015} within the {\sc reflex} interface \citep{Freudling13}. The two observations were barycentric corrected \citep[determined with {\sc astropy};][]{Astropy18}, continuum corrected by dividing by a low-order polynomial fit to the background continuum, and averaged together for the spectral analysis. 

\subsection{Spectral analysis}

The average spectrum obtained is shown in Fig.~\ref{figure:spectrum}, binned on 0.25\,nm (upper blue line). This is compared to HD~5689, an O6~V star as reported in \citet{hanson05} (lower black line). The identified lines are indicated on the figure. The H (Brackett series) and the He lines are the strongest features in the spectrum. The H lines are all in absorption, with no indication of emission lines that would be associated with an Oe/Be star.  The He\,{\sc ii} line at 2.1885$\micron$ is present, as found in O-type stars, while H\,{\sc i} lines are present indicating a type later than O3 \citep[e.g.][]{hanson05}. The relative ratio between the He\,{\sc ii} and He\,{\sc i} lines, at 1.6918\,$\micron$ and  1.7002\,$\micron$ respectively, is consistent with an O6 star. 
Upon comparison with the spectra presented in \citet{hanson05}, our best spectral fit points to an O6\,V star. The spectrum also shows N\,{\sc iii} (or possibly C\,{\sc iii} line) at $\sim2.116$\,$\micron$ in emission, as has been previously identified in O-type dwarf stars, as well as the unknown line at 1.650\,$\micron$ \citep{hanson96,hanson05}. The expected C\,{\sc iv} emission line at 2.078\,$\micron$
\citep{hanson05} is not seen but, due to the weakness of the line, it may have been lost in the correction for telluric absorption. In addition, an unknown line is observed at 1.567\,$\micron$, which is consistent with the spectrum of an O6~V star as reported in \citet{roman18}. We therefore conclude that \ircounter\ is an O6~V type star.

\begin{figure*}
	\includegraphics[width=\textwidth]{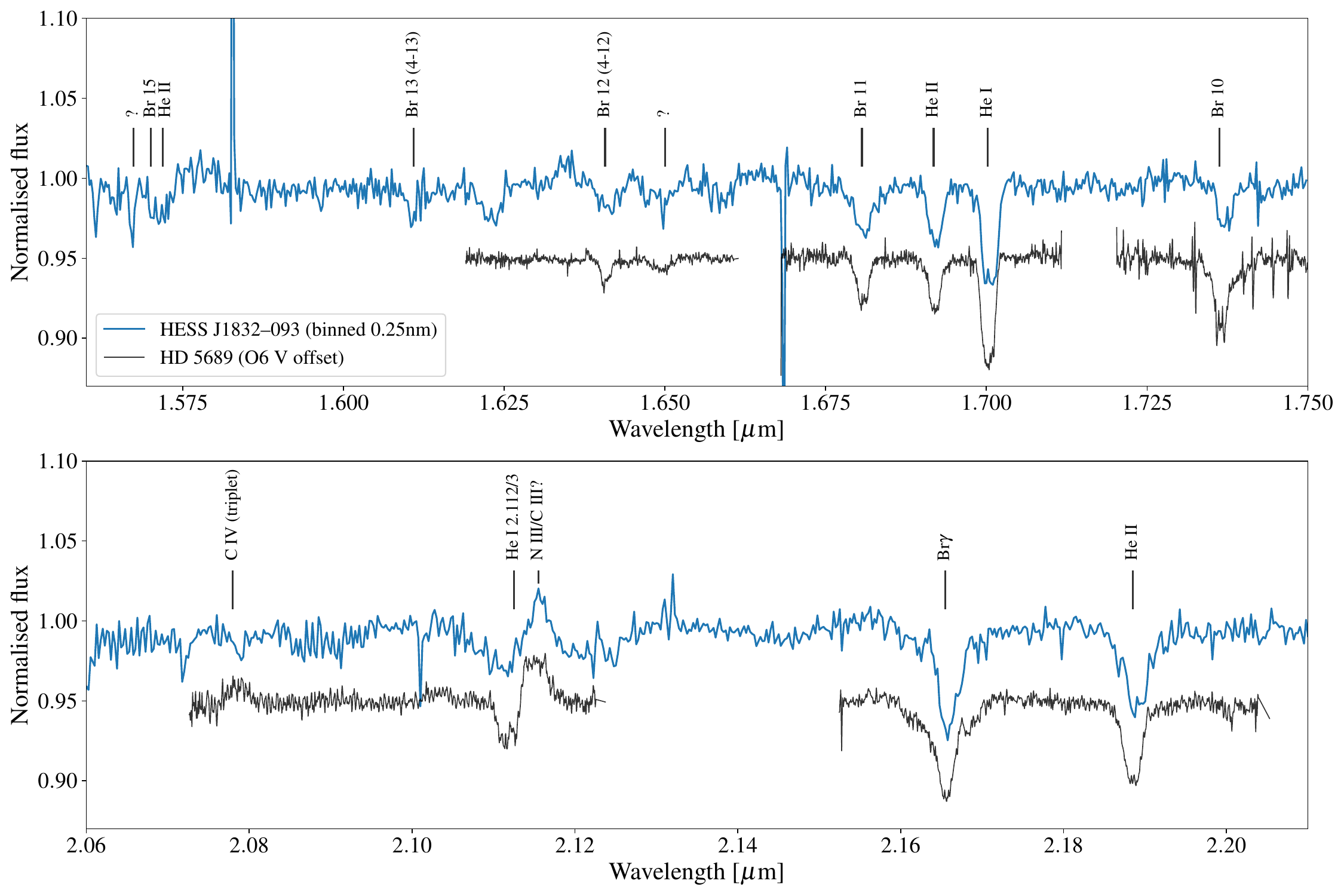}
 \caption{X-Shooter spectrum obtained in the analysis reported here, averaged over two observations, and binned on 0.25 nm (upper blue line). This is compared to HD~5689, an O6V star (lower black line, offset for clarity)  as reported by \citet{hanson05}. The identified H and He absorption lines are marked, as well as the N\,{\sc iii } (C\,{\sc iii}?) emission line. Also marked are the unidentified lines at 1.567\,$\micron$  \citep{roman18} and 1.650\,$\micron$ \citep{hanson05}, as well as the expected position of the C\,{\sc iv} emission line (see text for details).} 
  \label{figure:spectrum}
\end{figure*}

\subsection{Radial velocity search}

The two X-Shooter observations were taken at different times  in order to search for a change in the radial velocity of the source. For comparison to the phase folded \textit{Fermi}-LAT and \textit{Swift} light curves shown in fig.~8 in \citet{marti20}, we adopt  $T_0 = 54524.9979255$\,MJD and the orbital period of $P=87.016$\,d which places the two X-Shooter observations at orbital phases $\phi=0.67$ and $\phi =0.94$.\footnote{The correct value of the period found and used in figs.~7 \& 8 in \citet{marti20} is given in their figure captions, and is not the value given in the body of the text (Mart\'i-Devesa \& Reimer, private communication).} Note that the value of the orbital phase is very sensitive to the orbital period used.

We searched for a difference in the radial velocity between the two observations by fitting a Gaussian profile to several spectral lines. The noise on some individual lines prevented good-quality fits to the data, while for the cleaner cases we did not find a statistically significant difference between the mid-point. We note that the lack of a radial velocity difference in the two observations could also indicate that the intrinsic radial velocity is small, depending on the orbital geometry and the inclination angle of the binary system.  Additional observations will allow a cross-correlation between spectra to be performed, which can achieve a higher precision measurement of the radial velocities of \hessj.

%
\section{Discussion}
\label{section:discussion}

\subsection{Spectral type of the massive companion}

The X-shooter observations, obtained with a signal-to-noise of $>$100, clearly classify the proposed counterpart as an O6~V type star. This classification is consistent with the earlier ranges proposed by \citet{marti20} and \citet{Tam2020}. It is interesting to note that, with this spectral classification, the massive companions of all gamma-ray binaries that contain an O-type star have a similar spectral type. The massive companions in LS\,5039, 1FGL\,J1018.6$-$5856, LMP~P3, and 4FGL\,J1405.1$-$6119 are an ON6.5~V ((f)), O6~V ((f)), O5~III, and an O6.5~III star, respectively \citep{casares05,waisberg15,Seward12,Corbet19}. Therefore, to date, no gamma-ray binary has been detected with a spectral type later than O6.5 if the optical companion is not a Be star. All the known gamma-ray binaries containing Be stars also have a very similar spectral type of O9.5~Ve or B0~Ve \citep{negueruela11,casares05,aragona10, Camilo09}. However, this is not unexpected as gamma-ray binaries are likely precursors to more typical accretion-driven HMXBs \citep[e.g.][]{dubus17}, and Be X-ray binaries are known to show a peak at spectral type B0, which is linked to the evolution of these binary systems \citep[e.g.][]{negueruela02}. 

The reason for this apparent grouping around an O5/O6 spectral type is unclear. The initial mass function (IMF) predicts that the number of stars, $N$, decreases with increasing mass $M$ as $dN/d (\log M) \propto M^{-\Gamma}$, with $\Gamma \sim 1.35$ \citep[for $M\gtrsim 1$\,M$_\odot$;][]{salpeter55,Bastian2010}. Therefore, later O-type sources should be more common than more massive, earlier O-type sources. One possible reason for the grouping may be that to produce a gamma-ray binary the wind of the massive companion needs to have a high enough wind momentum to either produce a strong enough shock or efficiently confine the pulsar wind, in cases where a neutron star is powering the system. O-type stars display strong winds, which are powered by the intense radiation from these hot stars.  The mechanical momentum of their winds depends, therefore, on their luminosity, $L$, through the so-called wind momentum-luminosity relationship \citep[WLR;][]{Kudritzki00}, according to which $\log{D_{\rm mom}} \propto x \log{(L/L_{\sun})}$, where $D_{\rm mom} = \dot{M} v_\infty (R/R_{\sun} )^{0.5}$ is known as the modified stellar wind momentum. Here, $\dot{M}$ and $v_\infty$ are the mass-loss rate and velocity of the stellar wind, respectively, and $R$ is the stellar radius. 
A fit to O-type stars indicates that $x \sim [1.5 - 2.1$], with a possible indication of a break below $L/L_{\sun} \sim 10^5$ \citep[see e.g.][and references therein]{Kudritzki00, bjorklund21,marcolino22}.

Therefore, this may suggest that there is a preference for gamma-ray binaries to form with earlier, more luminous, O-type stars as they will have a higher wind momentum. However, since the IMF shows that the number of stars decreases with mass, spectral types earlier than O5 will be rare. This is illustrated in Fig.~\ref{figure:imf_wind}, where the shape of the IMF is shown in arbitrary units on the left axis, while the WLR \citep[using the fit from][]{bjorklund21} is shown on the right axis for O\,V and O\,III stars. Here the relation between stellar luminosity and stellar mass is estimated by interpolating between the values given in \citet{martins05}. This will introduce the observed grouping around a similar spectral type for systems containing O-type stars.  If this is correct, it suggests that gamma-ray binaries will only be found with later spectral type massive companions if they are Be stars, as this provides a denser wind in the circumstellar disc.

\begin{figure}
	\includegraphics[width=\columnwidth]{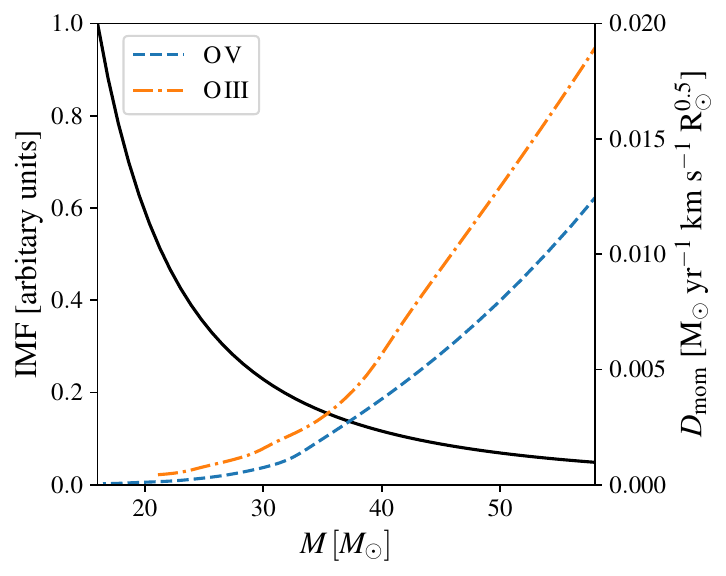}
 \caption{Sketch of the comparison of the IMF and the wind-momentum-luminosity relation for O-type stars. The IMF is in arbitrary units, assuming $\Gamma=1.35$ (black line; left axis), and the wind-momentum-luminosity relation (right axis) for O\,V (dashed blue line) and O\,III (dash-dotted orange line) type stars uses the fit from \citet{bjorklund21}, assuming the relation between mass and luminosity interpolated from \citet{martins05}. }
  \label{figure:imf_wind}
\end{figure}

\subsection{Distance to the source and lack of optical counterpart}

From the spectral type we can estimate the distance to the source from the distance modulus formula $m-M = -5 + 5 \log_{10}(d) + A_\lambda$. For an O6\,V type star we adopt the values from \citet{Martins06}, which gives $M_K = -4.13\pm0.15$ and $(J-K)_0 = -0.21$, where the error is estimated from the difference in magnitude for the immediate earlier and later type star, listed in the tabulated values. The observed colour for the NIR counterpart is $(J-K)_{\rm 2MASS} = 3.35\pm0.06$ which we correct via the formula outlined in \citet{Carpenter01} to the \citet{Bessell88} system, giving $(J-K) = 3.46 \pm 0.07$. This gives a colour excess of $E(J-K) = (J-K) - (J-K)_0 = 3.67\pm0.07$.  
We then calculate the extinction from $A_K / E(J-K) = 2.4 (\lambda_K / {\rm \mu m})^{-1.75}$ \citep{Draine89} where we use $\lambda_K=2.2\,{\rm \mu m}$, which gives an extinction of $A_K = 2.22\pm 0.04$. From this, and correcting the 2MASS $K_S$ magnitude to the \citet{Bessell88} system \citep{Carpenter01}, gives a distance to the source  of $d = 6.7 \pm 0.5$\,kpc, where the error is mainly due to the uncertainty in the absolute magnitude.
This distance is rather larger than the $4.4\pm0.4$\,kpc, previously assumed in e.g.\ \citet{marti20}. This would, however, only imply a luminosity $\sim 2$ times larger than previously considered, which would still imply that \hessj\ is a lower X-ray/gamma-ray luminosity gamma-ray binary \citep[see e.g.\ discussions in][and references therein]{Mori2017,Eger2016}. 

The colour excess in NIR also implies a large extinction at optical wavelengths. Using the interstellar extinction law from \citet{Rieke85} ($R=3.09\pm0.03)$ a rather high extinction in the V band of $A_V = 21.6\pm 1.4$ is derived. At a distance of 6.7\,kpc this would imply an apparent visual magnitude of $V = 31\pm1$,  \citep[assuming $M_V=-4.99$;][]{Martins06} much fainter than the current upper limits.  Similarly we estimated the expected magnitude in $I$, using  $M_I = -4.56\pm0.15$ \citep[based on $(I-V)_0=0.43$ in][]{Wegner94} and $A_I/A_V = 0.482$ \citep{Rieke85} which would suggest $I = 20.0 \pm 1.4$.  Given the upper limit established by the NOT observations (see \S\ref{sec:NOT}) is fainter than this, this suggests that the absorption in the direction of the source is higher than the standard Galactic extinction.

An estimate of the hydrogen column density in the direction of the source can also be made from $N_H/A_V = (1.79\pm0.03)\times10^{21}\,{\rm cm}^{-2}$ \citep{Predehl95}. From the value found for the visual extinction, this gives $N_H = (3.9 \pm 0.3)\times10^{22}\,{\rm cm}^{-2}$ , which is lower than the $N_H = 9.5\times10^{22}\,{\rm cm}^{-2}$, found from X-ray observations by \citet{Mori2017}. However, the estimates calculated here are for a standard Galactic extinction, which also implies that there is additional extinction in the direction of the source. Another possibility is that there is absorption which is intrinsic to the source, i.e.\ not due to interstellar absorption. However, to date this has not been observed for other gamma-ray binaries \citep[see e.g.][and references therein]{boschramon07,corbet16,an15,takahashi09}.

\section{Conclusions}

We report on new TJO, NOT, and X-Shooter observations in the NIR of the proposed counterpart to \hessj. From a comparison to the library of sources in \citet{hanson05}, and the relative ratios of the He\,{\sc i} and He\,{\sc ii} lines we classify the star as an O6\,V type. Based on this spectral type, we estimated the distance to the source to be $d = 6.7 \pm 0.5$\,kpc, but emphasise that the high value of the extinction found in this direction will affect this result. The high value of the extinction found, suggests that no optical observations will be possible of the massive counterpart, and all future studies (for example of the radial velocity of the companion star) will have to rely on observations in the near-infrared. This spectral classification further suggests a possible grouping around spectral type O5/O6, for gamma-ray binaries containing O-type stars. We suggest this grouping may be due to the interplay between the IMF and the WRL.

\section*{Acknowledgements}
Based on observations collected at the European Organisation for Astronomical Research in the Southern Hemisphere under ESO programmes 0104.D-0311(A)/0104.D-0311(B). Based on observations made with the Nordic Optical Telescope, owned in collaboration by the University of Turku and Aarhus University, and operated jointly by Aarhus University, the University of Turku and the University of Oslo, representing Denmark, Finland and Norway, the University of Iceland and Stockholm University at the Observatorio del Roque de los Muchachos, La Palma, Spain, of the Instituto de Astrofisica de Canarias. The data presented here were obtained in part with ALFOSC, which is provided by the Instituto de Astrofisica de Andalucia (IAA) under a joint agreement with the University of Copenhagen and NOT. The Joan Oró Telescope (TJO) of the Montsec Observatory (OdM) is owned by the Catalan Government and operated by the Institute for Space Studies of Catalonia (IEEC).
This research has made use of NASA’s Astrophysics Data System.  This research has made use of the SIMBAD database, operated at CDS, Strasbourg, France. This work made use of Astropy:\footnote{http://www.astropy.org} a community-developed core Python package and an ecosystem of tools and resources for astronomy. BvS acknowledges support by the National Research Foundation of South Africa (Grant Number 119430). PB and MR acknowledge the financial support from the State Agency for Research of the Spanish Ministry of Science and Innovation under grants PID2019-105510GB-C31/AEI/10.13039/501100011033, PID2019-104114RB-C33/AEI/10.13039/501100011033, and PID2022-138172NB-C43/AEI/10.13039/501100011033/ERDF/EU, and through the Unit of Excellence María de Maeztu 2020-2023 award to the Institute of Cosmos Sciences (CEX2019-000918-M). We acknowledge financial support from Departament de Recerca i Universitats of Generalitat de Catalunya through grant 2021SGR00679.  IN acknowledges the financial support of the Spanish Ministerio de Ciencia e Innovaci\'on (MCIN) with funding from the European Union NextGenerationEU and Generalitat Valenciana in the call Programa de Planes Complementarios de I+D+i (PRTR 2022), project HIAMAS (reference ASFAE/2022/017), as well as the Agencia Estatal de Investigaci\'on (MCIN/AEI/10.130~39/501~100~011~033/FEDER, UE) under grant PID2021-122397NB-C22. AP acknowledges funding from the INAF Research Grant “Uncovering the optical beat of the fastest magnetised neutron stars (FANS)” and from the Italian Ministry of University and Research (MUR),
PRIN 2020 (prot. 2020BRP57Z) “Gravitational and Electromagnetic-wave Sources in the Universe with current and next generation detectors (GEMS)''.

\section*{Data Availability}

The data presented here are available based on reasonable requests to the authors.



\bibliographystyle{mnras}
\bibliography{bibliography} 








\bsp	
\label{lastpage}
\end{document}